\documentclass[showpacs,prl,twocolumn,superscriptaddress]{revtex4}
\usepackage{epsfig}% Include figure files
\usepackage{bm}% bold math

\begin{document}

\title{Valence bond crystal in a pyrochlore antiferromagnet with orbital degeneracy}
\author{S. Di Matteo}
\affiliation{Laboratori Nazionali di Frascati INFN, via E. Fermi 40, C.P. 13, 
I-00044 Frascati (Roma) Italy}
\affiliation{Dipartimento di Fisica, Universit\`a di Roma III, via della Vasca
  Navale 84, I-00146 Roma Italy}
\author{G. Jackeli}
\altaffiliation[Also at ] {E. Andronikashvili Institute of Physics,
Georgian Academy of Sciences, Tbilisi, Georgia.} 
\affiliation{Institut Laue Langevin, B. P. 156, F-38042,
 Grenoble, France}
\author{C. Lacroix}
\affiliation{Laboratoire L. N\'eel, CNRS, BP 166, 38042 Grenoble Cedex 09, 
France}
\author{N. B. Perkins}
\affiliation{Laboratori Nazionali di Frascati INFN, via E. Fermi 40, C.P. 13, 
I-00044 Frascati (Roma) Italy}
\affiliation{Bogoliubov Laboratory of Theoretical Physics, JINR, 141980, Dubna, Russia}
\date{\today}
\begin{abstract}
We discuss the ground state of a pyrochlore lattice of 
threefold-orbitally-degenerate $S=1/2$ magnetic ions. 
We derive an effective spin-orbital Hamiltonian and 
show that the orbital degrees of freedom can modulate the spin
exchange, removing the infinite spin-degeneracy characteristic of 
pyrochlore structures. 
The resulting state is a collection of spin-singlet dimers, with a residual
degeneracy due to their relative orientation. This latter  is
lifted by a magneto-elastic interaction, 
induced in the spin-singlet phase-space, that forces a tetragonal
distortion. 
Such a theory provides an explanation for the 
helical spin-singlet pattern observed 
in the B-spinel MgTi$_2$O$_4$. 
\end{abstract}
\maketitle

Geometrically frustrated antiferromagnets have 
reached an increasing interest in the past decade \cite{1}.
The reason is that their ground states are 
highly degenerate and can evolve in a variety of ways:  
they can remain liquid down to the lowest 
temperatures due to quantum effects \cite{4}, or
lift their degeneracy via the 
order-out-of-disorder mechanism \cite{5}, or through a phase 
transition that lowers the local symmetry of the lattice \cite{ueda,tcher}. 
In this Letter we want to point out and 
discuss another scenario, that can appear
when magnetic ions of frustrated lattices possess also an orbital degeneracy. 
The physical behavior of such systems
may be drastically different from that of pure spin models, as the occurrence
of an orbital ordering can 
modulate the spin exchange, thus lifting the geometrical degeneracy of the underlying lattice.
In the following we focus on a system with 
threefold-orbitally-degenerate $S=1/2$ magnetic ions in a 
corner-sharing tetrahedral (pyrochlore) lattice.
This model is suitable to describe $d^1$-type transition-metal compounds, like the B-spinel MgTi$_2$O$_4$.
Here the magnetically active Ti$^{3+}$-ions form a pyrochlore lattice
and are characterized by one single electron in the threefold degenerate
$t_{2g}$-manifold.
MgTi$_2$O$_4$ undergoes a metal-to-insulator transition on cooling below 
260 K, with an
associated cubic-to-tetragonal lowering of the symmetry \cite{isobe}. At the transition the 
magnetic susceptibility continously  decreases and saturates, in the
insulating phase, to a value which is anomalously small for  spin $1/2$ local moments:
for this reason the insulating phase has been interpreted as a spin-singlet. 
Subsequent synchrotron and neutron powder diffraction
experiments have revealed that the low-temperature crystal
structure is made of alternating short and long Ti-Ti bonds
forming a helix about the tetragonal $c$-axis \cite{schmidt}.
These findings have suggested a removal of the pyrochlore degeneracy
by a one-dimensional (1D) helical dimerization of the spin pattern, 
with spin-singlets
located on short bonds. This phase can be regarded  as a 
valence bond crystal (VBC) since the long-range order of spin-singlets 
(dimers) extends throughout the whole pyrochlore lattice.

Here we describe the microscopic theory behind the stabilization of this VBC
ground state. 
We  argue that, remarkably,  such a  novel phase can be realized 
on a  pyrochlore lattice  because of orbital degeneracy, 
without invoking
any exotic interactions.
The existence of orbitally-driven 
VBC had been suggested for cubic lattice of $d^9$ compounds
 \cite{feiner}.
For $d^2$ compounds with frustrated lattices, 
the orbital order is shown to induce a spin-singlet
ground state, for triangular lattices \cite{pen},
or a spin ordered one, for pyrochlore lattices \cite{TS}. 
Yet, the peculiar case of $d^1$ spinel compounds leads to new results: 
the onset of an orbitally driven VBC state on the pyrochlore lattice.

{\it Effective Hamiltonian:} We first derive a superexchange spin-orbital 
Hamiltonian \`a la Kugel-Khomskii (KK) \cite{kugel} for threefold
orbitally-degenerate $d^1$-ions on a pyrochlore lattice.
We assume that the low-temperature insulating 
phase of MgTi$_2$O$_4$ is of Mott-Hubbard type.
We work in the cubic crystal class and look for possible instabilities towards symmetry reductions. 
Our parameters are the nearest-neighbor (NN) hopping term $t$,
 the Coulomb on-site repulsions $U_1$ (within the same 
orbital) and $U_2$ (among different orbitals), and the Hund's exchange, $J_H$.
For $t_{2g}$ wavefunctions the relation  $U_1=U_2+2J_H$ holds due to rotational
symmetry in real space.
The orbital occupancies of $t_{2g}$ orbitals, $n_{\alpha\beta}$ ($\alpha,\beta=x,y,z$), are expressed in terms of the 
pseudospin $\vec{\tau}=1$, with 
the correspondence: $\tau^{z}=-1\rightarrow|yz\rangle$, $\tau ^z=0\rightarrow |xy\rangle$, 
and $\tau ^z=1\rightarrow |xz\rangle$. 
At first, we consider only the leading part of the hopping term, due to the
largest $dd\sigma$ element, and discuss later the effects
of  smaller contributions (e.g. $dd\pi$).
The  $dd\sigma$ overlap  in $\alpha\beta$ plane 
connects only  the corresponding orbitals of the same $\alpha\beta$ type.
Thus, the total number of electrons in each orbital state is
a conserved quantity and the orbital part of the effective Hamiltonian 
$H_{\rm eff}$ is 
Ising-like:
\begin{eqnarray}
&&H_{\rm eff}=  
-J_1\sum_{\langle ij\rangle} {\big [}\vec S_i\cdot \vec S_{j} +3/4 {\big
  ]}O_{ij}
\label{spinorb}\\
&&+J_2\sum_{\langle ij\rangle} {\big [}\vec S_i\cdot \vec S_{j} -1/4 {\big ]}O_{ij}+
J_3\sum_{\langle ij\rangle} {\big [}\vec S_i\cdot \vec S_{j}-1/4 {\big ]}\tilde{O}_{ij}
\nonumber
\end{eqnarray}
where the sum is restricted to the NN  sites.
Introducing the projectors on the orbital states of site $i$,
$P_{i,xz}=\frac{1}{2}\tau_{iz}(1+\tau_{iz}) $, 
$P_{i,xy}=(1-\tau_{iz})(1+\tau_{iz})$ and 
$P_{i,yz}=-\frac{1}{2}\tau_{iz}(1-\tau_{iz})$, the orbital contributions along
the bond $ij$ in $\alpha\beta$-plane is given by $O_{ij}=P_{i,\alpha\beta}(1-P_{j,\alpha\beta})+P_{j,\alpha\beta}(1-P_{i,\alpha\beta})$ 
and $\tilde{O}_{ij}=P_{i,\alpha\beta}P_{j,\alpha\beta}$. 
The first and second terms in $H_{\rm eff}$ describe the 
ferromagnetic (FM)  $J_1=t^2/(U_2-J_H)$ and 
the antiferromagnetic (AFM) $J_2=t^2/(U_2+J_H)$ interactions,
respectively, and are active only when the two sites involved are occupied by 
different orbitals. The last term is AFM, with
$J_3=\frac{4}{3}t^2{\big [}2/(U_2+J_H)+1/(U_2+4J_H){\big ]}$, and 
is non-zero only when the two sites have the same orbital occupancy.
At this point it is useful to have an idea of the energy scales that play a
role in the Hamiltonian (\ref{spinorb}). We estimate 
 $t\equiv t_{\sigma} \simeq
0.32$ eV,
$J_H \simeq 0.64$ eV and $U_2\simeq 4.1$ eV \cite{mizo}.
Thus $\eta =J_H/U_2\simeq
0.15\ll 1$ and, just in order to present the
results in a more transparent form, we expand the exchange energies  around $\eta=0$.
We get $J_1\simeq J(1+\eta ) $, $J_2\simeq J(1-\eta )$ and $J_3\simeq
4J(1-2\eta )$ where $J=t^2/U_2 \simeq 25$ meV represents the overall energy
scale. In the following we measure all energies in units of $J$.

The main aspect of $H_{\rm eff}$ is that, due to $dd\sigma$-character of the hopping
terms, only some orbital configurations contribute to the energy: every
bond $ij$ in the $\alpha\beta$ plane has zero energy gain unless at least 
one of the two sites $i$ and $j$ has an occupied $\alpha\beta$ orbital. The strength, as well as the sign, of spin-exchange energy associated with
two NN sites $i$ and $j$ depends on their orbital occupations and the
direction of the $ij$ bond. 
The strongest bond
in the generic $\alpha\beta$-plane is characterized by both sites with $\alpha\beta$ occupancy: we shall call it $b_{0}$. Its exchange interaction is AFM and its spin Hamiltonian is given by:
%%%%%%%%%%%%%%%%%%%%%%%%%%%%%%%%%%%%%%%%%%%%%%%%
\begin{equation}
H_{b_0}=-1+2\eta+4[1-2\eta] \vec S_i\cdot \vec S_{j}.
\label{b0}
\end{equation}
%%%%%%%%%%%%%%%%%%%%%%%%%%%%%%%%%%%%%%%%%%%%%%%%%%%%
When the two sites of bond $ij$ in $\alpha\beta$-plane are occupied by one $\alpha\gamma$ and one
$\alpha\beta$ orbitals, $\gamma\not=\beta$ (bond $b_1$), one gets a weak FM
interaction:
%%%%%%%%%%%%%%%%%%%%%%%%%%%%%%%%%%%%%%%%%%%%%
\begin{equation}
H_{b_1}=-1-\eta/2-2\eta \vec S_i\cdot \vec S_{j}~.
\label{b1}
\end{equation}
%%%%%%%%%%%%%%%%%%%%%%%%%%%%%%%%%%%%%%%%%
Finally, the two sites of bond $ij$ in
$\alpha\beta$-plane
can be occupied by one $\alpha\gamma$ and one $\beta\gamma$ orbitals (bond  $b_2$), 
or, by two $\alpha\gamma$ (or $\beta\gamma$) orbitals (bond $b_3$). 
These bonds are noninteracting, as far as only $dd\sigma$ overlap is considered.

\begin{figure}
\epsfysize=27mm
\centerline{\epsffile{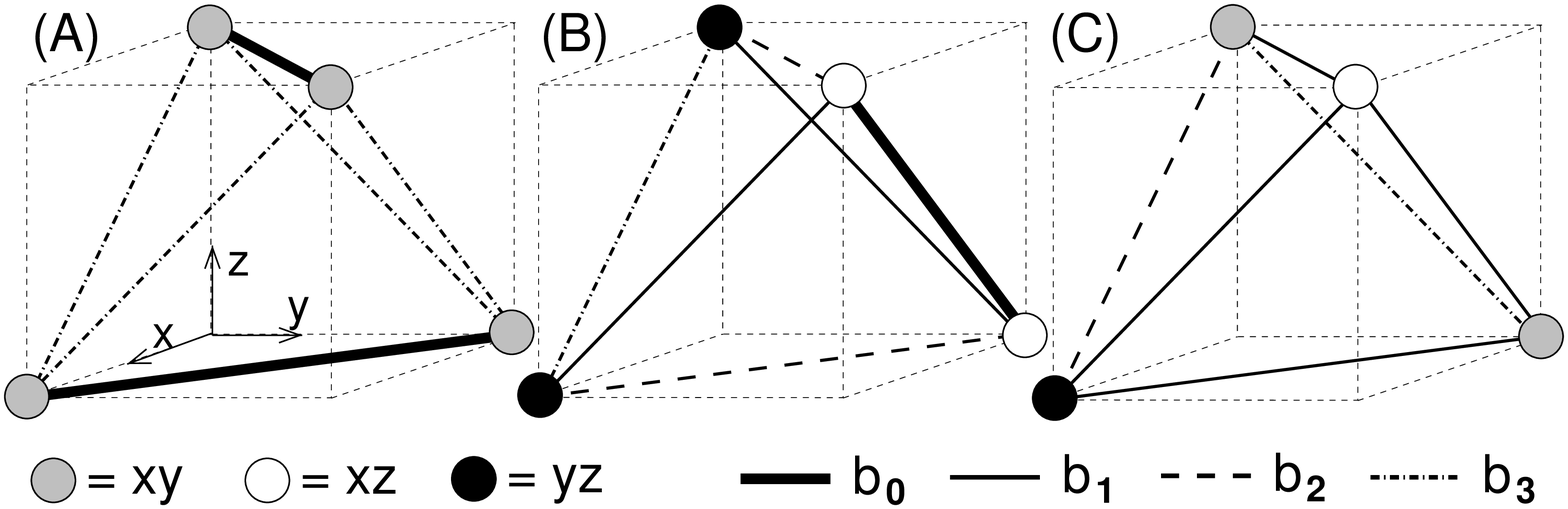}}
\caption{Orbital and bond arrangements on a tetrahedron for cases $A$), $B$) and $C$).}
\label{fig1}
\end{figure}

{\it Single tetrahedron}:
In one tetrahedron there are basically three possible orbital configurations 
to be considered
(see Fig. \ref{fig1}): {\it A}) all four sites have the same orbital occupancy 
(say $xy$) and thus only the two bonds in $xy$-plane (shown by solid lines in Fig. \ref{fig1}A) give a non-zero energy contribution;
 {\it B}) the two sites in one $\alpha\beta$-plane, e.g. $xz$, 
are both occupied by $xz$ orbitals,
 while at least one of the two sites in the other $xz$-plane is occupied by
 $xy$ or $yz$ orbitals (Fig. \ref{fig1}B) ; 
{\it C}) no bonds $ij$ in the plane $\alpha\beta$ of 
the tetrahedron is such as to have both sites occupied by the
$\alpha\beta$-orbital (Fig. \ref{fig1}C).
These three configurations are the bricks that allow to build the orbital pattern throughout
the whole pyrochlore lattice. Because of the Ising-form of orbital
interactions, in the following we can focus simply on these three cases, 
relying on the fact that configurations with a linear superposition of
orbitals on each site  must have a higher energy. 
We shall do only one exception to study a case with a
particular physical meaning, i.e., that of a "cubic" symmetry, where each site
is occupied by a linear superposition with equal weights 
of the three orbitals $\frac{1}{\sqrt{3}}[|xy\rangle+|xz\rangle+|yz\rangle]$, (case {\it D}). 

{\it Pyrochlore lattice:} Here we consider  possible coverings of the pyrochlore lattice through the various tetrahedra.

$A$) {\bf Heisenberg chains} -  When all tetrahedra of pyrochlore lattice are of
type $A$ (ferro-orbital ordering) then the effective Hamiltonian
(\ref{spinorb}) can be mapped into a set 
of one dimensional decoupled Heisenberg chains.
If, for example, all occupied orbitals are of $xy$-type, all chains
in $(1,\pm 1,0)$ cubic directions (see Fig. \ref{fig1}A) are decoupled.
The only interactions are due to AFM $b_0$-bonds described 
by the spin Hamiltonian (\ref{b0}).
Thus, the ground-state energy per site can be evaluated exactly by using the results for an Heisenberg 1D-chain \cite{1D}: $E_{A}=-2.77 (1-2\eta)$.
%%%%%%%%%%%%%%%%%%%%%%%%%%%%%%%%%%%%%%%%%%%%%%%%%%%%%%%%%%%%%%%%%%%%%%%%
%%%%%%%%%%%%%%%%%%%%%%%%%%%%%%%%%%%%%%%%%%%%%%%%%%%%%%%%%%%%%%%%%%%%%%%%

\begin{figure}
\epsfysize=38mm
\centerline{\epsffile{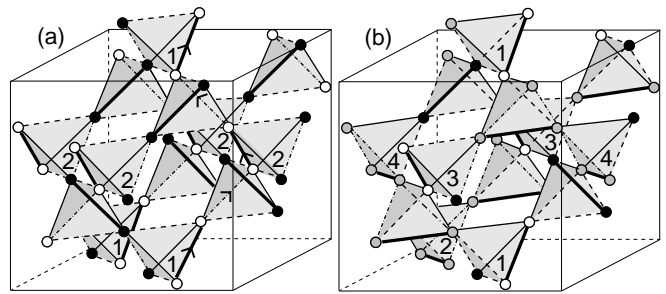}}
\caption{Two different coverings of the unit cubic cell through dimers,
  phase $B$. The same notations as in  Fig. \ref{fig1} are used. 
Locations of singlets are represented by thick links.
Different numbers correspond to inequivalent tetrahedra. 
{\it (a)} The experimental  phase of MgTi$_2$O$_4$:
  the helical dimerization pattern (indicated by arrows) 
is formed by alternating short $b_0$ and long  $b_3$ bonds.
All
  tetrahedra are in phase $B_1$; {\it (b)} a different
  covering of the cubic cell through all $B_i$
  tetrahedra. The inclusion of the magneto-elastic 
  coupling pushes these states higher in energy.}
\label{fig2}
\end{figure}

$B$) {\bf Dimer phase} - This state is made of $B$-type tetrahedra.
We distinguish three types of such tetrahedra.
All three are characterized by one strong $b_0$ bond, in a
$\alpha\beta$ plane. 
The other two ions in the opposite $\alpha\beta$-plane can either be occupied by two $\alpha\gamma$ ($\beta\gamma$)
orbitals, forming a $b_3$ bond, (case  $B_1$, 
the one shown in Fig. \ref{fig1}B),
or by one $\alpha\gamma$ and one
$\alpha\beta$ orbitals, linked in a $b_1$ bond, 
(case $B_2$), or, finally, by one $\alpha\gamma$ and one $\beta\gamma$
orbitals, forming a $b_2$ bond, (case $B_3$).
Since  $b_2$ and $b_3$ bonds do no 
contribute to the energy, 
this latter depends only on the number, $n_{b_0}$ and 
$n_{b_1}$, of $b_0$ and $b_1$ bonds in the unit cell.
As all three $B_i$ configurations are characterized by $n_{b_0}=1$ and 
$n_{b_1}=2$, all possible coverings of pyrochlore lattice by $B_i$ tetrahedra have the
same energy, even if ($n_{b_2}$, $n_{b_3}$) are different for
three $B_i$ tetrahedra ((2,1) for $B_1$; (1,2) for $B_2$; (3,0) for $B_3$). 
When pyrochlore lattice is covered by   $B_i$-types
tetrahedra (two possible  coverings  are shown in Fig. \ref{fig2}),
then  each spin is engaged in one strong AFM $b_0$ bond Eq. (\ref{b0}),  
and two weak FM
$b_1$-bonds Eq. (\ref{b1}).
Such coverings form a degenerate manifold and the corresponding energy can
be calculated as follows. 
In the limit $\eta \rightarrow 0$, the spin-only 
Hamiltonian can be solved exactly, as it can be decomposed into a sum 
of spin-uncoupled $b_0$ bonds. In this case the energy minimum is 
reached when the Heisenberg term of the $b_0$-bond is the lowest, i.e., 
for a pure quantum spin-singlet ($\vec S_i\cdot \vec S_{j}=-3/4$).
Remarkably, such spin-singlet (dimer) states, in the limit $\eta \rightarrow 0$, 
are also exact eigenstates of the full 
Hamiltonian (\ref{spinorb}).
As $\eta \ll 1$, the dimer state is stable against  
the weak FM interdimer interaction. In this case the magnetic  
 contribution along the FM $b_1$-bond is zero ($\langle \vec{S}_i\cdot \vec{S}_{j}\rangle=0$
for $i$ and $j$ belonging to different dimers)
  and we are led to an energy 
per site given by: 
$E_B=E_{b_0}/2+E_{b_1}=-3+\frac{7}{2}\eta$. Here $E_{b_{0(1)}}$ is the energy
of the bond $b_{0(1)}$. 

$C$) {\bf FM order}: Consider  the state where all tetrahedra are of type
$C$ (see Fig. \ref{fig1}C). There are four interacting $b_1$ bonds and two noninteracting
bonds ($b_2$ and $b_3$) per tetrahedron.
All non zero spin-exchanges are FM  given by Eq. (\ref{b1}).
The ground state for  this case is, thus, 
ferromagnetic with an energy $E_C=2E_{b_1}=-2(1+\eta)$.

$D$) {\bf Frustrated AFM}: 
The realization of this phase restores the full pyrochlore lattice symmetry, 
thus describing an ideal  cubic phase. 
All bonds are equivalent and by averaging Eq. (\ref{spinorb}) over the orbital
configurations
 on neighboring sites $i$ and $j$,
we obtain the spin Heisenberg  Hamiltonian on the
pyrochlore lattice:
$H_{D} = \sum_{\langle ij\rangle} (-5/9 +[4/9-16\eta/9]
\vec{S}_i\cdot \vec{S}_{j})$.
The system is thus highly frustrated and its ground state 
is a spin-liquid \cite{4}, whose energy per site is: $E_{D} \simeq -1.89 +0.89 \eta$.
Here we have used the ground-state energy estimated 
$(1/N)\sum_{ij} \vec{S}_i \cdot \vec{S}_{j}\simeq -0.5$ on the pyrochlore lattice \cite{canals}.

$E$) {\bf Mixed $A$ and $C$ configuration}: It 
is possible to cover the pyrochlore 
lattice also by means of a mixed configuration  with $A$ and $C$ types
tetrahedra. It can be visualized from Fig. \ref{fig2}b if, e.g., 
the four labeled tetrahedra would be  of A-type, with two
spin-singlets on strong $b_0$ bonds and three different orbital occupancy,
 and the four unlabeled tetrahedra of C-type  with no spin-singlets.
This dimer 
configuration is degenerate with the B-phase as far as only $dd\sigma$ 
overlap is considered, as, in average, $n_{b_0}=1$ and $n_{b_1}=2$.
Yet, in this case $n_{b_2}=1/2$ and $n_{b_3}=5/2$  and the degeneracy is
removed by $dd\pi$ overlap in favor of the B-phase (for which $n_{b_2}\geq
1$),  as the energy gain of the $b_2$ bond is $-t_{\pi}^2/[4t^2]$ (in units of
$J$), while that of the $b_3$ bond is $-t_{\pi}^2/[8t^2]$.

{\it Ground state manifold}:
 On the basis of the previous energy considerations, a simple phase diagram can be derived, in terms of $\eta$, the only free parameter available.
For $\eta =0$ the lowest ground-state energy is that of phase $B$.
With increasing $\eta$ we find only one phase transition
 at $\eta_c=2/11\simeq 0.18$, between dimer-phase $B$ and
FM phase $C$. As $\eta_c$ is above our estimated value of 
$\eta\simeq 0.15$, we can conclude that the ground state of MgTi$_2$O$_4$ is
described by the phase $B$ and is characterized 
by a frozen pattern of spin-singlets throughout the whole pyrochlore lattice that
removes the original spin degeneracy. 
Nonetheless, there is still a remaining  degeneracy 
to be lifted.
It is related to the freedom in the choice of the two orbitals on the 
tetrahedron bond opposite to the one of the singlet. 
Different choices of these orbitals give rise to inequivalent covering
patterns of the pyrochlore 
lattice with one dimer per tetrahedron  (see Fig. \ref{fig2}).
This degeneracy is given by the number of such dimer coverings and the
corresponding number of states can be estimated to grow with the
system size as ${\cal N} \sim 3^{N_T}=\sqrt{3}^{N}$ \cite{notedeg}. 
Here $N_T=N/2$ is the number of tetrahedra 
and we have ignored the contributions coming from closed loops (hexagons) on the pyrochlore
lattice.  This ground state manifold is different from a resonating valence bond
state, since each dimer covering  is frozen in an  exact eigenstate of the
Hamiltonian (\ref{spinorb}) for $\eta=0$. For finite $\eta$ the different dimer
patterns are not connected by Hamiltonian:  the bond corresponding to the dimer
in each tetrahedron is fixed, being determined by orbital pattern
and orbital degrees of freedom are Ising-like variables. 
Thus a tunneling between different dimer states cannot occur.

{\it Lifting of degeneracy}:
The above discussed degeneracy can not be removed within the KK-type model, not even introducing
$dd\pi$ and $dd\delta$ overlaps. The reason is related to the fact that the energy gain depends only 
on the total number of each type of bond ($n_{b_0}$, $n_{b_1}$, $n_{b_2}$, $n_{b_3}$) 
in the unit cell  and, 
in order to fill the whole crystal with a periodicity 
not lower than the one of the primitive cubic cell shown in Fig. \ref{fig2},
the average number of bonds $n_{b_i}$ per tetrahedron is the same,
whichever of the three building blocks $B_1$, $B_2$, or $B_3$ is used. 
It is given by $n_{b_0}=1$, $n_{b_1}=2$,
$n_{b_2}=2$, $n_{b_3}=1$, and it corresponds to the value of case $B_1$, that is 
the only one that allows to cover the whole cubic cell without mixing 
to other configurations (see Fig. \ref{fig2}a). Associated to the
$B_1$-phase, we have the  minimal cell enlarging (doubling instead of
quadrupling), and the maximal space subgroup (P$4_12_12$)
of the original face-centered Fd$\overline{3}$m cubic cell.

In order to substantiate these geometrical
 considerations, we need to find the physical mechanism that removes the $B$-manifold degeneracy in favor of the $B_1$-state. From the above discussion it follows that only correlations between bonds can lift it.
These correlations naturally appear if the magneto-elastic contribution to
 the energy is considered, as the orbitally-driven modulations of the 
spin exchange energies distort the underlying lattice through the spin-Peierls mechanism. 
In the degenerate $B$-manifold,
 every tetrahedron is characterized by a strong exchange on the bond $b_0$
 where the singlet is located. A reduction of the bond length enlarges the
 energy gain,
 because of the increase in the $dd\sigma$ overlap.
This selects the triplet-T deformation mode from the irreducible
 representations 
of the tetrahedron group, which is the only one that
 singles 
out one shorter bond \cite{note2}. This mode generates a tetragonal distortion
 of the tetrahedron, 
with short and long bonds located opposite to each other and four intermediate bonds.
Due to this mechanism, the position of the two $b_1$-bonds in the tetrahedron
 is uniquely determined:
 in order to maximize the superexchange energy gain by keeping the highest 
 value for $t_{\sigma}$, 
the intermediate-strength 
$b_1$-bonds are not allowed to lie on the long bond opposite to the singlet $b_0$. 
The elongation of the weak bonds of $b_{3}$ type is energetically more favorable.
It is possible to check that the only possibility to have such a constraint
for the whole cell is realized for the state $B_1$ (Fig. \ref{fig2}a). 
On the contrary, both cases $B_2$ and $B_3$ (Fig. \ref{fig2}b), 
do not allow to cover the cell
without at least one $b_1$-bond lying opposite to the singlet edge
 (e.g. tetrahedron 4 in Fig. 2b), thus with
an extra-energy cost. Hence, the energy is  minimized  
when all tetrahedra are of $B_1$ kind, with a $T$-type tetragonal distortion.
In this state all dimers are condensed in the ordered helical pattern shown in
Fig. \ref{fig2}a and form a VBC.
This dimerization pattern exactly reproduces the one
observed in the insulating phase of MgTi$_2$O$_4$ \cite{schmidt}. 
The present theory also predicts a peculiar orbital ordering in the dimerized
 phase: a ferro-type 
order along the helices 
 with antiferro-type order between them (see Fig. \ref{fig2}a).
This   orbital ordering can undergo an experimental 
test through Ti K edge natural circular
dichroism, which is sensitive to the chirality
of  $t_{2g}$ orbital order along 
the helix, when x-rays are shone along the helical axis.

Within the present scenario two transition temperatures are expected. 
The highest $T_{c_1}$, determined by the exchange coupling within the singlet 
($T_{c_1}\sim J_{3}\simeq 4J \simeq 1000 K$), corresponds to the transition from a paramagnet to a spin gap (dimer) state, with a ferro-orbital order on each dimer. This state can be regarded as a weakly-interacting gas of dimers and is highly degenerate
with respect to the dimers orientation.
The lowest transition temperature, given by the magneto-elastic coupling,
lifts the degeneracy through the spin-Peierls distortion.
At this temperature  dimers condense  and form the VBC
shown in Fig.\ref{fig2}a. The entropy involved in this
transition is estimated to be $\sim \ln [{\cal N}]/N=\ln \sqrt{3}$.
In the case of MgTi$_2$O$_4$, it is known \cite{isobe} that, 
with increasing temperature, this compound goes from an insulating to a metallic
phase at $T_{c_2} \simeq 260 K$. The transition to the metallic state
rules out the possibility of a high-temperature spin-singlet state with disordered
dimers and does not allow to evaluate the order of magnitude of the
magneto-elastic coupling. We can just estimate its lower limit as about
1/4$(\simeq T_{c_2}/T_{c_1})$ of the singlet-binding energy.  

In summary, we have derived a mechanism that allows to lift the geometrical
 degeneracy of a pyrochlore lattice of threefold-orbitally-degenerate
 $t_{2g}$, spin-1/2 magnetic ions like Ti$^{3+}$. We have singled out two
 relevant energy scales that govern its behavior: the main one is the
 superexchange energy gain,
 that drives the system into a spin-singlet dimer state with peculiar orbital pattern.  
The residual orientational degeneracy is then lifted through the
magneto-elastic interaction that optimizes the previous energy gain 
by distorting the bonds in the suitable directions and leading to a tetragonal distortion. 
This generates a condensate of dimers in a VBC state, forming
1D dimerized helical chains running
around the tetragonal $c$-axis, the one actually observed in MgTi$_2$O$_4$.
%%%%%%%%%%%%%%%%%%%%%%%%%%%%


\begin{thebibliography}{99}
\bibitem{notedeg} When a singlet is located on a given tetrahedron then each
  neighboring tetrahedron is left with only three possible choices for a
  singlet location.
\bibitem{note2} For a simple spin 
Hamiltonian it is the
doublet $E$-mode that removes the local degeneracy of 
the tetrahedron \cite{ueda}.
\end{thebibliography}
\end{document}